\documentclass[9pt]{elife}

\usepackage{lipsum} 
\usepackage[version=4]{mhchem}
\usepackage{siunitx}
\DeclareSIUnit\Molar{M}

\usepackage{listings}
\usepackage{xcolor}

\definecolor{codeorange}{rgb}{0.9,0.4,0.1}
\definecolor{codegreen}{rgb}{0,0.6,0}
\definecolor{codegray}{rgb}{0.5,0.5,0.5}
\definecolor{codepurple}{rgb}{0.58,0,0.82}
\definecolor{backcolour}{rgb}{0.95,0.95,0.93}

\lstdefinestyle{python_style}{
    backgroundcolor=\color{backcolour},   
    commentstyle=\color{codegreen},
    keywordstyle=\color{codeorange},
    numberstyle=\tiny\color{codegray},
    stringstyle=\color{codepurple},
    basicstyle=\ttfamily\small,
    breakatwhitespace=false,         
    breaklines=true,                 
    captionpos=b,                    
    keepspaces=true,                 
    numbers=left,                    
    numbersep=5pt,                  
    showspaces=false,                
    showstringspaces=false,
    showtabs=false,                  
    tabsize=2
}

\title{PyRates - A Code-Generation Tool for Dynamical Systems Modeling}

\author[1*]{Richard Gast}
\author[2]{Thomas R. Kn\"osche}
\author[1]{Ann Kennedy}
\affil[1]{Feinberg School of Medicine, Northwestern University, Chicago, US}
\affil[2]{Max Planck Institute for Human Cognitive and Brain Sciences, Leipzig, Germany}

\corr{richard.gast@northwestern.edu}{RG}

\begin{document}

\maketitle

\begin{abstract}
Mathematical models allow us to gain a deeper understanding of real-world dynamical systems.
One of the most powerful mathematical frameworks for modeling real-world phenomena are systems of differential equations.
In the vast majority of fields that use differential equations to investigate dynamical systems, numerical methods are essential for conducting model-based research.
Although many software solutions are available for the numerical study of differential equation systems, manually translating models between software environments is a time-consuming and highly error-prone process.
This lack of a common framework for implementing differential equation systems hinders progress in dynamical systems research and limits the shareability and reproducibility of results.

\textit{PyRates} is a Python-based software for modeling and analyzing dynamical systems.
It provides a user-friendly interface for defining models, which is based on a graph-based, hierarchical structure that mirrors the modular organization of real-world dynamical systems.
This design allows users to leverage the hierarchical structure of their systems and create their models with minimal effort.

Importantly, the core of \textit{PyRates} is a versatile code-generation system, which can translate user-defined models into "backend" implementations in various languages, including \textit{Python}, \textit{Fortran}, and \textit{Julia}.
This allows users to access a wide range of analysis methods for dynamical systems, eliminating the need for manual translation between code bases.
\textit{PyRates}'s code-generation system is also designed to be easily extended to support other programming languages and backends should they become relevant to dynamical systems research. 

We demonstrate \textit{PyRates}'s capabilities in three use cases, where it generates \textit{NumPy} code for numerical simulations via \textit{SciPy}, \textit{Fortran} code for bifurcation analysis and parameter continuations via \textit{PyCoBi}, and \textit{PyTorch} code for neural network optimization via \textit{RectiPy}.
Finally, \textit{PyRates} can be used as a model definition interface for the creation of new dynamical systems tools.
We developed two such software packages, \textit{PyCoBi} and \textit{RectiPy}, as extensions of \textit{PyRates} for specific dynamical systems modeling applications.
\end{abstract}


\section{Introduction}
\label{sec:intro}

Scientists have been using ordinary differential equation (ODE) systems to study real-world dynamical systems since the formulation of classical mechanics by Newton \citep{newton_philosophiae_1833,hubbard_differential_2013,strogatz_nonlinear_2018,hutt_synergetics_2020}. 
Disciplines as diverse as physics, biology, neuroscience, and earth sciences have applied ODE systems to model phenomena such as fluid dynamics, population growth, neural synchronization, and climate change.
While some simple ODE systems have analytical solutions, most real-world systems are too complex to study analytically \citep{strogatz_nonlinear_2018}.
Hence, numerical methods are critical to gain a scientific understanding of ODE systems \citep{stuart_dynamical_1998,meijer_hil_g_e_numerical_2009}.
For example, numerical methods can find solutions to complex problems such as the prediction of weather changes \cite{coiffier_fundamentals_2011}, the conditions for an ecosystem to approach a stable state \cite{jost_testing_2000}, or the optimal application of electrical stimulation for treating a neurological disorder \cite{west_stimulating_2022}.
Solving these problems using numerical methods can involve the integration of ODE systems with thousands of state variables, the application of automated parameter optimization algorithms in high-dimensional parameter spaces, or the automated detection of stable solutions of differential equation systems.

The research community has developed many software packages that efficiently implement the most widely used numerical analyses for dynamical systems (see Tab.~\ref{tab:ds_tools} for examples).
However, there is no standardization in how dynamical systems models must be formulated or how analysis models are implemented across different packages.
Additionally, different packages differ in their degree and style of software documentation, versioning, and automated testing.
These idiosyncrasies impede the adoption, reproducibility, shareability, and transparency of numerical dynamical system analysis results \citep{freire_computational_2012,topalidou_long_2015,gruning_practical_2018}.

Here, we present \textit{PyRates}, an open-source \textit{Python} toolbox for dynamical systems modeling.
\textit{PyRates} provides a flexible model definition language, which is parsed by the library's code-generation tools into output code that can be run in various third-party software packages or "backends".
The model definition language enables users to define simple mathematical operators (ordinary differential equations, ODEs) and connect them hierarchically to form networks of interacting elements.
Models defined via \textit{PyRates} can be translated for processing by any of its various backends (see Tab.~\ref{tab:ds_tools} for examples).
For example, the same model definition can be used to perform parameter optimization via the \textit{Julia} toolbox \textit{BlackBoxOptim.jl} \citep{feldt_blackboxoptimjl_2022} and bifurcation analysis via the \textit{Fortran}-based software \textit{Auto-07p} \citep{doedel_auto-07p:_2007}.
Thus, \textit{PyRates} offers (i) a simplified process for implementing dynamical system models with minimal potential for errors, (ii) a powerful model definition language that permits sharing of ODE implementations across different dynamical system analysis packages, and (iii) access to a wide range of dynamical system analysis methods through its code generation approach.


\subsection{Overview}
\textit{PyRates} was previously introduced as a toolbox for neural network modeling \citep{gast_pyratespython_2019}.
Its code generation capabilities have since grown considerably, and it has evolved into a general modeling framework for analyzing and translating dynamical systems.
The purpose of this paper is to demonstrate these capabilities, and introduce \textit{PyRates} as a code generation tool for dynamical systems modeling in general.

In the following sections, we will first compare \textit{PyRates} to other, related dynamical systems modeling software.
We will then present the software structure of \textit{PyRates} in detail, noting novel features that have been added since our previous manuscript \citep{gast_pyratespython_2019}.
This is followed by use cases that demonstrate the main features of \textit{PyRates} using a number of well-known dynamical system models.
Finally, we will discuss the limitations of the software, as well as its potential to foster dynamical systems modeling research.

\begin{table}[tbhp]
    \centering
    \begin{tabular}{|c|c|c|} \hline
        \textbf{Name} & \textbf{Description} & \textbf{Language}\\ \hline
        \parbox{5cm}{\textit{DifferentialEquations.jl} \\ \citep{rackauckas_differentialequationsjl_2017}} & \parbox{6cm}{\vspace{.5\baselineskip}Toolbox for numerical analysis of various types of differential equation systems\vspace{.5\baselineskip}} & \textit{Julia}\\ \hline
        \parbox{5cm}{\textit{BlackBoxOptim.jl} \\ \citep{feldt_blackboxoptimjl_2022}} & \parbox{6cm}{\vspace{.5\baselineskip}Toolbox for model-independent parameter optimization\vspace{.5\baselineskip}} & \textit{Julia}\\ \hline
        \parbox{5cm}{\textit{Auto-07p} \\ \citep{doedel_auto-07p:_2007}} & \parbox{6cm}{\vspace{.5\baselineskip}Toolbox for numerical parameter continuation and bifurcation analysis of ODE systems\vspace{.5\baselineskip}} & \textit{Fortran}\\ \hline
        \parbox{5cm}{\textit{SciPy} \\ \citep{virtanen_scipy_2020}} & \parbox{6cm}{\vspace{.5\baselineskip}Toolbox that includes methods for ODE integration and parameter optimization\vspace{.5\baselineskip}} & \textit{Python}\\ \hline
        \parbox{5cm}{\textit{PyTorch} \\ \citep{paszke2019pytorch}} & \parbox{6cm}{\vspace{.5\baselineskip}A machine learning library that includes methods for gradient-based parameter optimization\vspace{.5\baselineskip}} & \textit{Python}\\ \hline
        \parbox{5cm}{\textit{pygpc} \\ \citep{weise_pygpc_2020}} & \parbox{6cm}{\vspace{.5\baselineskip}Model-independent sensitivity and uncertainty analysis toolbox\vspace{.5\baselineskip}} & \textit{Python}\\ \hline
        \parbox{5cm}{\textit{DDE-BIFTOOL} \\ \citep{sieber_dde-biftool_2016}} & \parbox{6cm}{\vspace{.5\baselineskip}Toolbox for numerical parameter continuation and bifurcation analysis of delayed differential equation systems\vspace{.5\baselineskip}} & \textit{Matlab}\\ \hline
    \end{tabular}
    \vspace{.5\baselineskip}
    \caption{Exemplary list of dynamical systems analysis software packages.}
    \label{tab:ds_tools}
\end{table}


\subsection{Related Work}

As shown in section \ref{sec:vdp}, \textit{PyRates} supports numerical integration of differential equation systems and parallelized parameter sweeps.
While this feature is useful for model validation and small dynamical system analyses, it is not the main purpose of the software.
Other tools such as \textit{DifferentialEquations.jl} for \textit{Julia} \citep{rackauckas_differentialequationsjl_2017}, \textit{SciPy} for \textit{Python} \citep{virtanen_scipy_2020}, or \textit{XPPAUT} for \textit{Matlab} \citep{ermentrout_simulating_2003} offer a wide range of numerical ODE solvers. 
The main advantage of \textit{PyRates} is that it allows users to interface these tools from a single model definition, giving them the flexibility to choose the best solver for their purposes.

\textit{PyRates}'s code-generation approach sets it apart from dynamical system modeling frameworks such as COMSOL MultiPhysics \citep{multiphysics_introduction_1998}, \textit{PyDS} \citep{clewley_hybrid_2012}, \textit{PySD} \citep{houghton_advanced_2015}, \textit{Simupy} \citep{margolis_simupy_2017}, \textit{The Virtual Brain} \citep{sanz_leon_virtual_2013,sanz-leon_mathematical_2015}, the \textit{Brain Dynamics Toolbox} \citep{heitmann_brain_2018}, or the Brain Modeling Toolkit \citep{dai_brain_2020}, which provide a range of dynamical system analysis methods within a single framework.
These tools can be useful for minimizing implementation errors, and for users that want a single tool with a set of analysis and visualization options.
However, if a specific analysis method or algorithm is not provided, these tools lack the flexibility to interface with third-party software.
In contrast, \textit{PyRates}'s code-generation approach allows users to choose the best algorithms and implementations for each step in a dynamical system analysis pipeline.
For example, given a single model definition, \textit{PyRates} can export one piece of code to use \lstinline{scipy.optimize} to fit your model to data, another to use \textit{DDE-BIFTOOL} for bifurcation analysis around the optimized parameter set, and finally a third to generate time series in different parameter regimes via \textit{DifferentialEquations.jl}.

This code-generation framework makes \textit{PyRates} similar to tools such as \textit{Brian} \citep{goodman_brian_2009}, \textit{ANNarchy} \citep{vitay_annarchy:_2015}, \textit{RateML} \citep{van_der_vlag_rateml_2022}, \textit{NESTML} \citep{plotnikov_nestml_2016}, or \textit{NeuroML} \citep{kotter_towards_2001}.
All of these tools generate code from user-defined model equations and are designed for numerical integration of neurodynamic models.
Their use of code generation allows users to design a custom model via the software frontend, and obtain optimized code for backend implementation that is efficient on specific hardware.
However, they each only generate code for a specific third-party backend (such as C or \textit{Python}), and the generated code is not directly accessible to the user.
\textit{PyRates}, on the other hand, provides inherent access to the code generated for its different backends, while still offering run-time optimization options such as vectorizing the model equations or using function decorators like \textit{Numba} \citep{lam_numba_2015} (see the gallery example on run-time optimization at \url{https://pyrates.readthedocs.io/en/latest/}).
The user can easily manipulate \textit{PyRates}-generated code, for example to embed it into other scripts, thus maintaining full control even after the model is translated into a specific backend.
This is an advantage over other string-based code generation methods, as the generated code can be easily observed and analyzed.
Therefore, \textit{PyRates} is attractive to experts and scholars in dynamical system modeling.
It provides the flexibility to implement complex models and use expert-level analysis tools, while also offering full control over the model equations, allowing scholars to examine and adjust the output of \textit{PyRates} and gain a deeper understanding of the models and analysis techniques.

In summary, \textit{PyRates} is more than just an ODE solver.
It is a dynamical system modeling framework that offers a range of ODE solving options, but mostly stands out for (i) a simple, yet powerful model definition language (see section \ref{sec:frontend}), and (ii) translating these models into equation files for interfacing with other dynamical system tools (see section \ref{sec:backend}).


\section{Software Structure}

\textit{PyRates} is freely available on \textit{GitHub} and comes with detailed documentation.
Each version is released on \textit{PyPI} and can easily be installed using the \textit{pip} package manager.
For installation instructions, see the \textit{GitHub} repository: \url{https://github.com/pyrates-neuroscience/PyRates}.
The repository also provides information on officially supported \textit{Python} versions and the status of the extensive test library included with \textit{PyRates}.
The latter ensures that all main features and models are working as expected in the current version of \textit{PyRates}.
The structure of the software can be viewed in the API section of our documentation website: \url{https://pyrates.readthedocs.io/en/latest/}.

\begin{figure}[ht!]
    \centering
    \includegraphics[width=1.0\textwidth]{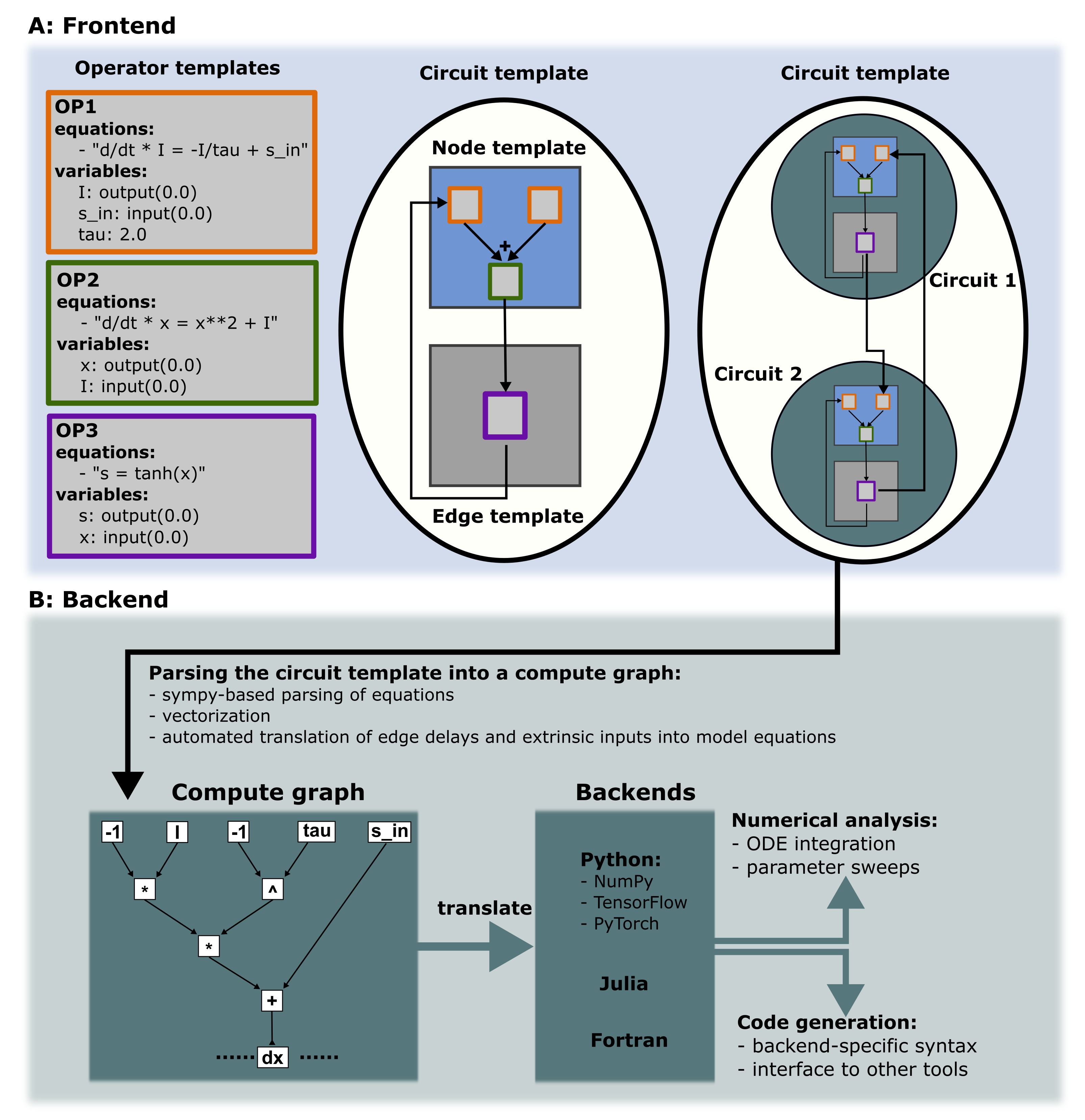}
    \caption{\textit{PyRates} software structure. \textbf{(A)} Depiction of the user interface: \textit{PyRates} models are implemented via different templates that can be defined via a \textit{YAML} or \textit{Python} interface. \lstinline{OperatorTemplate} instances are used to define equations and variables and serve as basic building blocks for \lstinline{NodeTemplate} and \lstinline{EdgeTemplate} instances. The latter can be used to define \lstinline{CircuitTemplate} instances which are used to represent the final models in \textit{PyRates}. \lstinline{CircuitTemplate} instances can also be incorporated in higher-level \lstinline{CircuitTemplate} instances to allow for complex hierarchies, as depicted by the coupling of \textit{circuit 1} and \textit{circuit 2} within a \lstinline{CircuitTemplate}. \textbf{(B)} Structure of the backend: Each model is translated into a compute graph, which in turn is parsed into a backend-specific model implementation. The latter can be used for code generation and numerical analyses.}
    \label{fig:pyrates_structure}
\end{figure}

\textit{PyRates} consists of a frontend and a backend.
The frontend provides a user-friendly interface for model definition, numerical simulations, and code generation, while the backend allows efficient evaluation of the model equations using a number of powerful programming languages and toolboxes.
See Fig.~\ref{fig:pyrates_structure} for a visualization of this structure.

\subsection{The PyRates Frontend \label{sec:frontend}}

Fig.~\ref{fig:pyrates_structure}A illustrates how models are defined via a hierarchy of template classes.
A detailed description of this template-based user interface is provided in our previous work \citep{gast_pyratespython_2019}, as well as in our online documentation (\url{https://pyrates.readthedocs.io/en/latest/}).

In short, \textit{PyRates} allows the implementation of dynamical system models of the form 

\begin{equation}
    \dot{\mathbf{y}} = \mathbf{F}(\mathbf{y}(t), \mathbf{\theta}, t, \mathbf{y}(t - \tau_1), ..., \mathbf{y}(t- \tau_n)), \label{eq:des}
\end{equation}

with $N$-dimensional state-vector $\mathbf{y}$ and $N$-dimensional vector-field $\mathbf{F}$.
This vector field can depend on the current state of the system $\mathbf{y}(t)$ as well as previous states of the system $\mathbf{y}(t - \tau_i) \forall i \in {1,...,n}$, a parameter vector $\mathbf{\theta}$ and time $t$.
Thus, \textit{PyRates} supports the implementation of autonomous and non-autonomous dynamical systems, and allows for the use of ordinary and delayed first-order differential equation systems.
For more information on the mathematical framework and syntax supported by \textit{PyRates}, see \url{https://pyrates.readthedocs.io/en/latest/math_syntax.html}.

The basic functional unit of PyRates is the \textbf{operator template}, which is composed of a differential equation of the form \eqref{eq:des} and its associated input, output, and intrinsic variables.
Operator templates are in turn organized into nodes and edges, where nodes represent the atomic units of the dynamical system, and edges represent coupling functions between these units.
Thus, the operators are combined such that they represent the underlying equations of the atomic units (nodes) and their connections (edges).
The final dynamical system is always created via a \textbf{circuit template}, which are defined by a set of nodes and their connecting edges.
One of the advantages of the template-based model definition is that each template can be used multiple times.
This is shown in Fig.~\ref{fig:pyrates_structure}A, where three operator templates are used multiple times in the nodes and edges in the circuit template.
Likewise, the nodes (edges) that share a structure in Fig.~\ref{fig:pyrates_structure}A only require a single node (edge) template for their definition.
Finally, as seen in Fig.~\ref{fig:pyrates_structure}A, the template-based model definition interface allows for the use of circuit templates to define higher-level circuits, enabling the creation of models of complex, hierarchically structured dynamical system.
For further documentation of the template user interface, see \url{https://pyrates.readthedocs.io/en/latest/template_specification.html}.


\subsection{The PyRates Backend \label{sec:backend}}

Fig.~\ref{fig:pyrates_structure}B illustrates the working principles of the \textit{PyRates} backend.
Whenever a model template is used for simulations or code generation, the model is first translated into a compute graph.
This is done using the equation parsing functionalities of \textit{SymPy}, a well-known \textit{Python} library for symbolic mathematics \citep{meurer_sympy_2017}.
The resulting graph represents all variables and the mathematical operations connecting them, creating a flow chart from the ODE system input to its output, i.e. the vector field of the model.
Following its construction, the compute graph is translated into a backend-specific function for the evaluation of the vector field.
This function can be used directly for numerical simulations of the system dynamics, or it can be written to a file, with the syntax and file type depending on the chosen backend.
Currently, the following backends are available in \textit{PyRates}:
\begin{itemize}
    \item \textit{NumPy} \citep{harris_array_2020},
    \item \textit{TensorFlow} \citep{abadi2016tensorflow},
    \item \textit{PyTorch} \citep{paszke2019pytorch},
    \item \textit{Fortran 90} \citep{adams_fortran_1993},
    \item \textit{Julia} \citep{bezanson_julia_2017},
    \item \textit{Matlab} \citep{higham_matlab_2016}.
\end{itemize}
Due to the modular structure and open-source nature of \textit{PyRates}, additional backends can be added with relatively little effort.
Generated function files can be used to interface other tools such as the ones listed in Tab.~\ref{tab:ds_tools}, or numerical integration of the model equations or parameter sweeps can directly be performed in \textit{PyRates}.
In that case, \textit{PyRates} will automatically use the generated function file.


\subsection{PyRates as a Model Definition Interface}

\textit{PyRates} can also be used as a model definition interface for more specialized dynamical systems tools.
Tools that extend \textit{PyRates} can take advantage of its template-based, hierarchical model definition system, and use \textit{PyRates}'s code generation capacities to translate model definitions for a target backend.
Here, we present two \textit{Python} tools that we developed using \textit{PyRates} as their model definition interface: \textit{PyCoBi}, for parameter continuation and bifurcation analysis, and \textit{RectiPy} for recurrent neural network modeling.
Both tools are part of the collection of open-source software provided with \textit{PyRates} and are freely available at \url{https://github.com/pyrates-neuroscience}.

\subsubsection*{\textit{PyCoBi}} This package provides specialized support for parameter continuation and bifurcation analysis, two common numerical computing tasks in the characterization of dynamical systems.
\textit{PyCoBi} is based on the \textit{Fortran} software \textit{Auto-07p}, one of the most popular and powerful tools for parameter continuations.
By leveraging the code generation functionality of \textit{PyRates}, \textit{PyCoBi} provides a modern user interface to \textit{Auto-07p} that does not require any \textit{Fortran} coding (although users can also use \textit{PyCoBi} on existing Fortran files.) 
We demonstrate the functionality of \textit{PyCoBi} in section \ref{sec:qif}, where we use \textit{PyRates} to generate \textit{Fortran} files for a dynamical system and use those files to perform bifurcation analysis via \textit{PyCoBi}.

\subsubsection*{\textit{RectiPy}} This package extends \textit{PyRates} with custom methods for recurrent neural network optimization and simulation.
\textit{RectiPy} uses \textit{PyRates} both to define networks of recurrent rate or spiking neurons and to translate those networks into a \textit{PyTorch} graph.
It further provides high-level routines for gradient-based parameter optimization and numerical integration of the network equations via the \textit{PyTorch} graph.
We demonstrate the functionalities of \textit{RectiPy} and how it integrates \textit{PyRates} as a user interface in section \ref{sec:li}.


\section{Use Examples}

In this section, we demonstrate different stages of the \textit{PyRates} workflow using well-known ODE models.
We show how different dynamical system analysis methods can be applied to these models via \textit{PyRates}, and demonstrate the flexibility that \textit{PyRates} offers in analyzing dynamical system model dynamics and parameter dependencies.

The dynamical system models used in the examples below come pre-implemented with \textit{PyRates} and are explained in detail in our online documentation at \url{https://pyrates.readthedocs.io/en/latest/}, while scripts to reproduce the results and figures of each of our use examples at \url{https://www.github.com/pyrates-neuroscience/use_examples}.
Custom dynamical system model implementation in \textit{PyRates} is also described in the online documentation.


\subsection{Using \textit{PyRates} for Numerical Simulations and Parameter Sweeps \label{sec:vdp}}

In this example, we demonstrate how \textit{PyRates} can be used to perform numerical simulations and parameter sweeps.
We study a Van der Pol oscillator receiving periodic input from a simple Kuramoto oscillator, and examine its entrainment to the Kuramoto oscillator frequency as a function of the input strength and frequency.
This is done using a \textit{PyRates} function that performs multiple, vectorized numerical integrations of the ODE system (\ref{eq:vdp1}-\ref{eq:vdp2}), one for each parametrization of interest.
The equations of the ODE system are:

\begin{align}
    \dot x &= z, \label{eq:vdp1}\\
    \dot z &= \mu z (1 - x^2) - x - J sin(2\pi \theta), \\
    \dot \theta &= \omega. \label{eq:vdp2}
\end{align}

The state variables of the ODE system (\ref{eq:vdp1} - \ref{eq:vdp2}) are the Van der Pol oscillator state variables $x$ and $z$ and the Kuramoto oscillator phase $\theta$, and the system parameters are given by the damping constant $\mu$, the input strength $J$, and the intrinsic frequency of the Kuramoto oscillator $\omega$.

Equations for both the Van der Pol and Kuramoto oscillators are pre-implemented in \textit{PyRates}.
For comprehensive reviews of the properties of these oscillators, see \citep{acebron_kuramoto_2005,kanamaru_van_2007}.
The following code uses the \lstinline{NodeTemplate} class to load the definitions of the Van der Pol oscillator and Kuramoto oscillator, then uses the \lstinline{CircuitTemplate} class to define the network of nodes and edges that make up the dynamical system given by equations (\ref{eq:vdp1} - \ref{eq:vdp2}):

\begin{lstlisting}[language=Python, caption=Definition of the Van der Pol oscillator model.]
from PyRates import CircuitTemplate, NodeTemplate

# define nodes
VPO = NodeTemplate.from_yaml(
    "model_templates.coupled_oscillators.vanderpol.vdp_pop"
    )
KO = NodeTemplate.from_yaml(
    "model_templates.coupled_oscillators.kuramoto.sin_pop"
    )

# define network
net = CircuitTemplate(
    name="VPO_forced", nodes={'VPO': VPO, 'KO': KO}, 
    edges=[('KO/sin_op/s', 'VPO/vdp_op/inp', None, {'weight': 1.0})]
    )
\end{lstlisting}

With the model loaded into \textit{PyRates}, we can use numerical integration to generate time series of its dynamics for different values for $J$ and $\omega$.
To minimize the runtime of this problem, we use the function \lstinline{pyrates.grid_search}, which takes a set of multiple model parametrizations and performs the numerical integration in a single combined model by vectorizing the model equations.
The code below defines a parameter sweep with 20 values of $J$ and 20 values of $\omega$, resulting in $N = 400$ model parametrizations.

\begin{lstlisting}[language=Python, caption=Parameter sweep over periodic forcing parameters in the Van der Pol oscillator model., label={lst:grid_search}]
# imports 
import numpy as np
from PyRates import grid_search

# define parameter sweep
n_om = 20
n_J = 20
omegas = np.linspace(0.3, 0.5, num=n_om)
weights = np.linspace(0.0, 2.0, num=n_J)

# map sweep parameters to network parameters
params = {'omega': omegas, 'J': weights}
param_map = {'omega': {'vars': ['phase_op/omega'], 
                       'nodes': ['KO']},
             'J': {'vars': ['weight'], 
                   'edges': [('KO/sin_op/s', 'VPO/vdp_op/inp')]}
            }

# perform parameter sweep
results, res_map = grid_search(
    circuit_template=net, param_grid=params, param_map=param_map,
    simulation_time=T, step_size=dt, solver='scipy', method='DOP853',
    outputs={'VPO': 'VPO/vdp_op/x', 'KO': 'KO/phase_op/theta'},
    inputs=None, vectorize=True, clear=False, file_name='vpo_forced',
    permute_grid=True, cutoff=cutoff, sampling_step_size=dts
    )
\end{lstlisting}

The \lstinline{grid_search} call takes the given \lstinline{circuit_template} and creates copies of it for each set of parameters in \lstinline{param_grid}. 
It adjusts the parameters of each copy accordingly, using the information in \lstinline{param_map} to locate the parameters that should be adjusted.
It then places all copies of the \lstinline{circuit_template} in one big model and performs the simulation, with the remaining arguments controlling the numerical integration procedure.
Given a set of $N$ different model parametrizations, this procedure results in an implementation of the ODE system (\ref{eq:vdp1} - \ref{eq:vdp2}) where each variable in the equations is represented by a vector of length $N$.
For the numerical integration, we instructed \lstinline{grid_search} to use a Runge-Kutta algorithm of order 8 with automated adaptation of the integration step size, available through the \lstinline{scipy.integrate.solve_ivp} method of \textit{SciPy} \citep{virtanen_scipy_2020}.
Alternative choices of numerical integration methods are available via the keyword arguments \lstinline{solver} and \lstinline{method} of \lstinline{grid_search}.

\begin{figure}[ht!]
    \centering
    \includegraphics[width=1.0\textwidth]{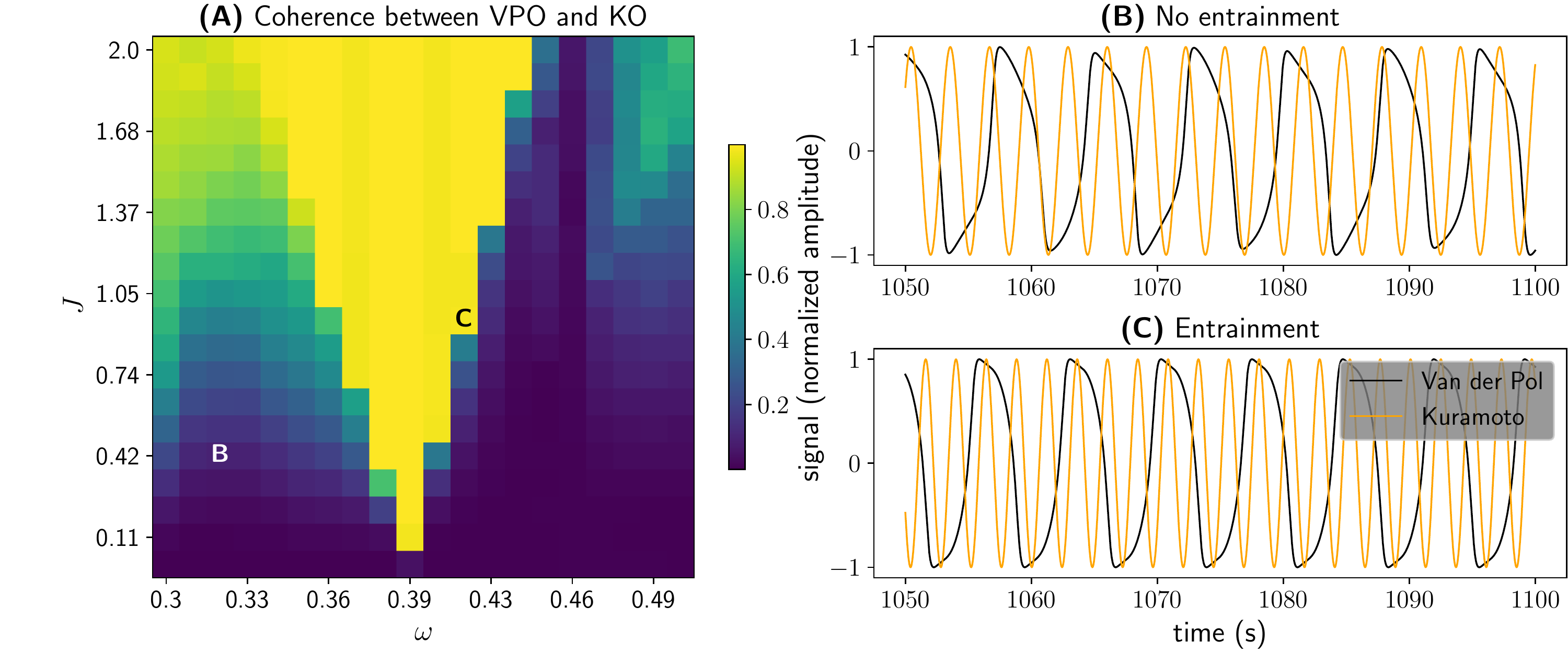}
    \caption{Entrainment of the Van der Pol oscillator in response to periodic forcing. \textbf{(A)} Coherence between the state variables $x$ of the Van der Pol oscillator (VPO) and $\theta$ of the Kuramoto oscillar (KO). For each pair of $\omega$ and $J$, we bandpass-filtered $x$ at the frequency $\omega$ and extracted the phase of the bandpass-filtered signal via the Hilbert transform. We then created a sinusoidal signal from the VPO and KO phases and used \lstinline{scipy.signal.coherence} to calculate the coherence between the two sinusoids. The result is depicted as color-coding. \textbf{(B and C)} State variables $x$ (black) and $\theta$ (orange) displayed over time. \textbf{(B)} No entrainment of the VPO phase for $\omega = 0.33$ and $J = 0.5$. \textbf{(C)} Entrainment of the VPO phase to the KO phase for $\omega = 0.42$ and $J = 1.0$.}
    \label{fig:vdp}
\end{figure}

We use the returned values of the \lstinline{grid-search} call in Listing \ref{lst:grid_search} to compute the coherence between $\theta$ and $x$ for each set of $\omega$ and $J$.
This results in a triangularly shaped coherence profile, also know as an Arnold tongue (see Fig.~\ref{fig:vdp}A), which describes the characteristic entrainment behavior of a non-linear oscillator subject to a periodic driving force \citep{boyland_bifurcations_1986}.

The larger the difference between the driving frequency and the intrinsic frequency of the non-linear oscillator (or one of its harmonics), the stronger the required amplitude of the driving signal for entraining the oscillator to the driving frequency.
Our example confirms that the Van der Pol oscillator expresses this behavior.
In Fig.~\ref{fig:vdp}B, we show an example where the driving force $J$ was too small to entrain the oscillator given the substantial difference between $\omega$ and the intrinsic frequency of the oscillator.
In Fig.~\ref{fig:vdp}C, on the other hand, we show an example where the driving force $J$ was sufficiently high and the difference between $\omega$ and the intrinsic frequency of the oscillator was sufficiently low to entrain the oscillator.

Thus, the above example demonstrates how \textit{PyRates} can be used to perform the first steps of any dynamical system analysis, numerical integration of the differential equation system and parameter sweeps.
We studied the entrainment of the Van der Pol oscillator to the Kuramoto oscillator frequency as a function of the input strength and frequency.
We used the \textit{PyRates} function \lstinline{pyrates.grid_search} to perform multiple, vectorized numerical integrations of the ODE system.
The results confirm previous findings on the entrainment of a non-linear oscillator and show that the numerical integration and parameter sweep functionalities of \textit{PyRates} work as expected.


\subsection{Using PyRates for bifurcation analysis \label{sec:qif}}

In this example we present \textit{PyCoBi}, one of \textit{PyRates}'s extensions for applying numerical bifurcation analysis to a dynamical system model.
Numerical bifurcation analysis is an essential tool to study qualitative changes in model dynamics caused by small variations in model parametrization \citep{meijer_hil_g_e_numerical_2009,kuznetsov_elements_2013}.

We study the neurodynamic model described in detail in \citep{gast_mean-field_2020}: A mean-field model of coupled quadratic integrate-and-fire (QIF) neurons with spike-frequency-adaptation (SFA):

\begin{align}
    \tau \dot r &= \frac{\Delta}{\pi \tau} +2 r v, \label{eq:qif1}\\
    \tau \dot v &= v^2 +\bar\eta + I(t) - a + J r \tau -(\pi r \tau)^2,\\
    \tau_a \dot a &= x, \\
    \tau_a \dot x &= \alpha \tau_a r - 2x - a. \label{eq:qif2}
\end{align}

The state variables of this model are $r$ and $v$, the average firing rate and membrane potential of the QIF population, and $a$ and $x$, which describe the SFA dynamics.
For more details on the model equations and constants, see \citep{gast_mean-field_2020}.
We are interested in the effects of the SFA strength $\alpha$ and the average neural excitability $\bar\eta$ on population dynamics, and would use \textit{Auto-07p} \citep{doedel_auto-07p:_2007} to carry out the bifurcation analysis.
The goal is to reproduce the bifurcation diagrams reported in \citep{gast_mean-field_2020}, where the effects of $\alpha$ and $\bar\eta$ on the dynamics of (\ref{eq:qif1} - \ref{eq:qif2}) have already been investigated.

\textit{Auto-07p} requires used-supplied \textit{Fortran} files that include the model equations and constants.
As demonstrated below, \textit{PyRates} can be used to generate these files.
First, we need to load the model into \textit{PyRates}.
Since the dynamical system given by (\ref{eq:qif1} - \ref{eq:qif2}) exists as a pre-implemented model in \textit{PyRates}, this can be done via a single function call:

\begin{lstlisting}[language=Python, caption=Definition of the QIF model.]
from PyRates import CircuitTemplate
qif = CircuitTemplate.from_yaml(
    "model_templates.neural_mass_models.qif.qif_sfa"
    )
\end{lstlisting}

After the model is loaded, it can be used to generate the input required for \textit{Auto-07p}:

\begin{lstlisting}[language=Python, caption=Auto-07p file generation via PyRates.]
qif.get_run_func(func_name='qif_run', file_name='qif_sfa', step_size=1e-4,
                 backend='fortran', solver='scipy', vectorize=False, 
                 float_precision='float64', auto=True)
\end{lstlisting}

This method generates two files required to run \textit{Auto-07p}: a \textit{Fortran 90} file containing the model equations and a simple text file containing the meta parameters of \textit{Auto-07p}.
The equation file, which includes a vector field evaluation function named \lstinline{func_name}, can be found in the location indicated by \lstinline{file_name}; the meta parameter file, named \lstinline{c.ivp}, is in the same directory.
Providing the keyword argument \lstinline{auto=True}, ensures that the output files are in a format compatible with \textit{Auto-07p}.

At this point, \textit{PyRates} has generated the meta parameters and equation files needed for parameter continuations and bifurcation analysis in \textit{Auto-07p}.
To demonstrate this, we use \textit{PyCoBi}, which allows the calling of \textit{Auto-07p} functions from \textit{Python}.
In the example below, we perform a simple numerical integration of the ODE system over time, allowing it to converge to a steady-state solution that we can then further analyze via parameter continuations.
For the example to execute without errors, provide a path to the installation directory of  \textit{Auto-07p} via \lstinline{auto_dir=<path>}.

\begin{lstlisting}[language=Python, caption=Numerical integration of the QIF model via \textit{Auto-07p}.]
# initialize PyCoBi
from pycobi import ODESystem
qif_auto = ODESystem(working_dir=None, auto_dir=<path>, init_cont=False)

# perform numerical integration
t_sols, t_cont = qif_auto.run(
    e='qif_sfa', c='ivp', name='time', DS=1e-4, DSMIN=1e-10, EPSL=1e-08,
    EPSU=1e-08, EPSS=1e-06, DSMAX=1e-2, NMX=1000, UZR={14: 5.0}, STOP={'UZ1'}
    )
\end{lstlisting}

The arguments provided to \lstinline{ODESystem.run} are mostly identical to the arguments required to run \textit{Auto-07p}, which are explained in detail in the documentation at: \url{https://github.com/auto-07p/auto-07p/tree/master/doc}.
Most importantly, pointers to the generated equation and meta parameters files have been provided via the arguments \lstinline{e='qif_sfa'} and \lstinline{c='ivp'}, respectively.

\begin{figure}[ht!]
    \centering
    \includegraphics[width=1.0\textwidth]{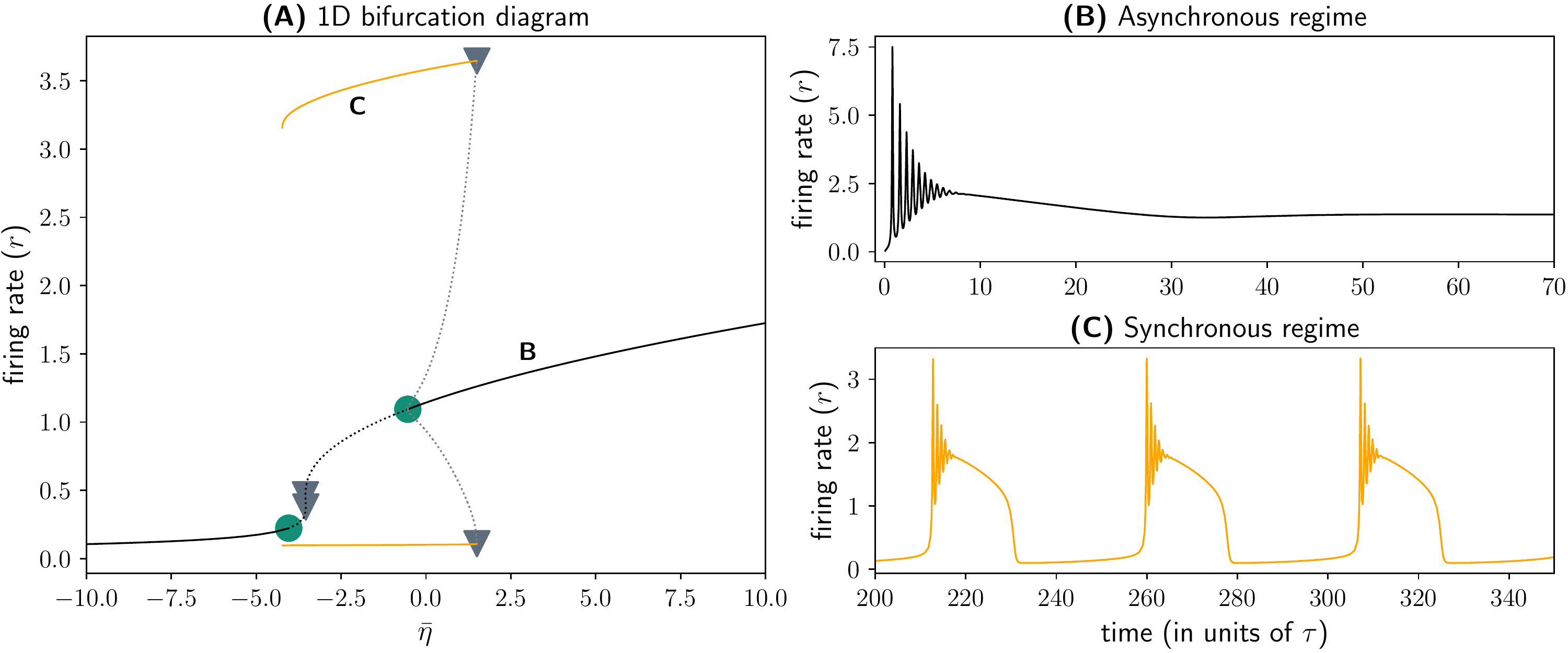}
    \caption{Bifurcation analysis of the QIF model. \textbf{(A)} Bifurcation diagram showing the solutions of Eqs.(\ref{eq:qif1} - \ref{eq:qif2}) in the state variable $r$ as a function of the parameter $\bar\eta$. Solid (dotted) lines represent stable (unstable) solutions. Bifurcation points are depicted as symbols along the solution branches. Green circles represent Hopf bifurcations whereas grey triangles represent fold bifurcations. \textbf{(B)} Convergence of the average firing rate $r$ of the QIF model to a steady-state solution in the asynchronous, high-activity regime ($\bar\eta = 3$). \textbf{(C)} Convergence of the average firing rate $r$ of the QIF model to a periodic solution in the synchronous, oscillatory regime ($\bar\eta = -2$).}
    \label{fig:qif}
\end{figure}

In Fig.~\ref{fig:qif}B, we see that the QIF mean-field model converged to a steady-state solution within the provided integration time of the ODE system (\ref{eq:qif1} - \ref{eq:qif2}).

Starting from this steady-state solution, we can perform parameter continuations and automated bifurcation analysis\citep{meijer_hil_g_e_numerical_2009} via \textit{PyCoBi}.
We first continued the steady-state solution we calculated previously in the background input parameter $\bar\eta$.
As can be seen in Fig.~\ref{fig:qif}A, the steady-state solution branch undergoes a number of bifurcations within the examined range of $\bar\eta$: Two fold bifurcations and two sub-critical Hopf bifurcations.
By continuing the unstable periodic solutions emerging from the latter, we next identified fold of limit cycle bifurcations that give rise to a regime of synchronized oscillations (see Fig.~\ref{fig:qif}C).

These results confirm the findings reported in \citep{gast_mean-field_2020}, where a more detailed description of the QIF model's bifurcation structure is provided.
Thus, we have successfully demonstrated that \textit{PyRates} provides an interface to the parameter and bifurcation analysis software \textit{Auto-07p}, one of the most powerful tools for studying solutions of differential equation systems and how they change with underlying system parameters.


\subsection{Parameter fitting in a delay-coupled leaky integrator model \label{sec:li}}

In this example, we demonstrate the capacity of \textit{PyRates} as a model definition interface for other tools and show how it can support the definition of large-scale, delay-coupled dynamical systems. 
To do so, we show how \textit{RectiPy} leverages \textit{PyRates} as a frontend and allows \textit{PyRates} models to be optimized using any \textit{PyTorch} parameter optimization routine.
In the example below, we use \textit{RectiPy} for gradient-based parameter optimization in a recurrent neural network model with delay coupling that is implemented in \textit{PyRates}.

\subsubsection{Building the network with \textit{RectiPy}}

We use a set of $N$ leaky integrators with non-linear delay-coupling as an exemplary model:

\begin{align}
    \dot u_i = - \frac{u_i}{\tau} + I_{ext}(t) + k \sum_{j=1}^N J_{ij} \tanh(\Gamma_{ij} * u_j),\label{eq:li1}\\ 
    \Gamma_{ij}(t) = \frac{a_{ij}^{b_{ij}} t^{b_{ij}-1} e^{a_{ij}t}}{(b_{ij}-1)!}, \label{eq:li2} 
\end{align}

where $\tau$ is a global decay time constant, $k$ is a global coupling constant, $J_{ij}$ are connection-specific coupling strengths, and $I_{ext}$ is a variable that allows for extrinsic forcing.
The term $\Gamma_{ij} * u_j$ is a convolution of the rate $u_j$ with the gamma kernel given by eq.~\eqref{eq:li2}.
This type of gamma-kernel convolution is a popular model for delay-coupled systems with distributed delays \citep{smith_distributed_2011, gast_role_2021}.

The following code implements a network of $N = 5$ coupled leaky integrators using \textit{RectiPy}'s \lstinline{Network} class, with random coupling weights and gamma kernel parameters as given by eqs.~(\ref{eq:li1}, \ref{eq:li2}).

\begin{lstlisting}[language=Python, caption=Initialization of the delay-coupled leaky integrator model in \textit{RectiPy}., label=lst:rectipy]
import numpy as np
from rectipy import Network

# network parameters
node = "neuron_model_templates.rate_neurons.leaky_integrator.tanh_pop"
N = 5
J = np.random.uniform(low=-1.0, high=1.0, size=(N, N))
D = np.random.choice([1.0, 3.0], size=(N, N))
S = D*0.3
dt = 1e-3

# initialize network
net = Network(dt=dt, device="cpu")

# add a recurrently coupled population of leaky integrators to the network
net.add_diffeq_node(
    "tanh", node=node, weights=J, edge_attr={'delay': D, 'spread': S}, 
    source_var="tanh_op/r", target_var="li_op/r_in", 
    input_var="li_op/I_ext", output_var="li_op/u"
    )

\end{lstlisting}

As shown in line 16 of Listing \ref{lst:rectipy}, \lstinline{rectipy.Network.add_diffeq_node} provides an interface for adding a \textit{PyRates} model as a node to a \lstinline{rectipy.Network} instance.
\lstinline{Network.add_diffeq_node} first uses \lstinline{NodeTemplate.from_yaml(node)} to set up the governing equations of each network node.
It then uses the connectivity weights provided via the \lstinline{weights} keyword argument together with all additional edge attributes (\lstinline{edge_attr}) to create a \lstinline{CircuitTemplate}, and fill it with $N^2$ edges.
This is implemented by a call to \newline \lstinline{pyrates.CircuitTemplate.add_edges_from_matrix}, a method that adds edges of the following form to the network:

\begin{lstlisting}[language=Python, caption=Definition of an edge in \textit{PyRates} .]
edge = ("<pi>/tanh_op/m", "<pj>/li_op/m_in", None, 
        {"weight": C_ij, "delay": D_ij, "spread": S_ij}
       )
\end{lstlisting}

Here, $D_{ij}$ and $S_{ij}$ refer to the mean and variance of the gamma kernel $\Gamma_{ij}$ and are related to its parameters via $D_{ij} = \frac{a_{ij}}{b_{ij}}$ and $S_{ij} = \frac{a_{ij}}{b_{ij}^2}$.
Each edge definition that includes both the \lstinline{"delay"} and the \lstinline{"spread"} keyword is automatically translated into a gamma kernel convolution of the source variable by \textit{PyRates}.
\textit{PyRates} implements the convolution operation as a set of coupled ODEs that it adds to the model, using the 'linear chain trick' \citep{smith_distributed_2011}.

All string-based keyword arguments provided in lines 17-19 of Listing \ref{lst:rectipy} are pointers to model variables defined in the \textit{YAML} template specified in line 5 of Listing \ref{lst:rectipy}.
This ensures that the network equations generated by \textit{PyRates} are properly integrated into the \textit{PyTorch} graph.
For example, the keyword argument \lstinline{input_var="li_op/I_ext"} indicates that any input provided to the network should enter the network equations via the variable \lstinline{I_ext} that is defined in the operator \lstinline{li_op} of the \lstinline{NodeTemplate}. 

Having constructed the network, \textit{RectiPy} uses the \textit{PyTorch} backend of \textit{PyRates} to generate a vector field function that can be used for simulations and parameter optimization via \textit{PyTorch}.
This way, \textit{RectiPy} extends \textit{PyRates} to enable quick generation of \textit{PyTorch} compute graphs from \textit{YAML} templates.
\textit{RectiPy} can thus provide a powerful user interface for simulating and fitting recurrent neural networks with minimal coding effort.

\subsubsection{Performing parameter optimization in a \textit{RectiPy} model}

We next demonstrate the use of \textit{RectiPy} for parameter optimization. Our goal will be to recover the values of model parameters in the \lstinline{rectipy.Network} instance defined in the previous section, specifically the global time constant $\tau = 2.0$ and the global coupling constant $k = 1.0$. To do so, we'll sample the model's response to a $200 Hz$ sinusoidal driving input, and then use this observed response to fit the values of $k$ and $\tau$ in a second, identical model instance in which $k$ and $\tau$ are initialized from a uniform distribution over $[0.1, 10.0]$.

We can sample the target model's activity similarly to in \textit{PyTorch}:

\begin{lstlisting}[language=Python, caption=Generation of the target signal for parameter optimization in \textit{RectiPy}., label=lst:targets]
# simulation parameters
dt = 1e-3
steps = 30000
f = 0.2
beta = 0.1

# simulate target signal
targets = []
for step in range(steps):
    I_ext = np.sin(2*np.pi*freq*step*dt) * beta
    u = net.forward(I_ext)
    targets.append(u)
\end{lstlisting}

Where, as in \lstinline{torch.nn}, \lstinline{rectipy.Network.forward} generates the output variable $u$ from the input $I_{ext}$, using the functional relationship defined by eqs.\eqref{eq:li1} and \eqref{eq:li2}.
Alternatively, numerical simulation can be performed in a single line with:

\begin{lstlisting}
 obs = net.run(inputs, sampling_steps=1)   
\end{lstlisting}

where \lstinline{inputs} is a vector of the extrinsic input to the network at each time point.
This \lstinline{rectipy.Network.run} method returns an instance of \lstinline{rectipy.Observer}, which provides access to all network state variables recorded during the simulation.

Next, we will fit our second network model to the observed dynamics of our target network. The code example below shows how to perform a single optimization step in \textit{PyTorch} using a mean-squared error loss function and the resilient backpropagation algorithm \citep{riedmiller_direct_1993} to calculate the gradient of the error with respect to the free parameters $\tau$ and $k$.

\begin{lstlisting}[language=Python, caption=Parameter optimization step in \textit{RectiPy}., label=lst:opt]
import torch

# loss function definition
loss = torch.nn.MSELoss()

# optimizer definition
opt = torch.optim.Rprop(net.parameters(), lr=0.01)

# calculate cumulative error over entire target signal
mse = torch.zeros(1)
for step in range(steps):
    I_ext = np.sin(2*np.pi*f*step*dt) * beta
    u_target = targets[step]
    u = net.forward(I_ext)
    mse += loss(u, u_target)

# optimization step
opt.zero_grad()
error.backward()
opt.step()
    
\end{lstlisting}

A target signal can be fitted by iterating over optimization steps until convergence.
Alternatively, the entire optimization procedure is also available via the \lstinline{rectipy.Network.fit_bptt} method:

\begin{lstlisting}
 obs = net.fit_bptt(inputs, targets, optimizer="rprop", loss="mse", lr=0.01)  
\end{lstlisting}

\begin{figure}[ht!]
    \centering
    \includegraphics[width=1.0\textwidth]{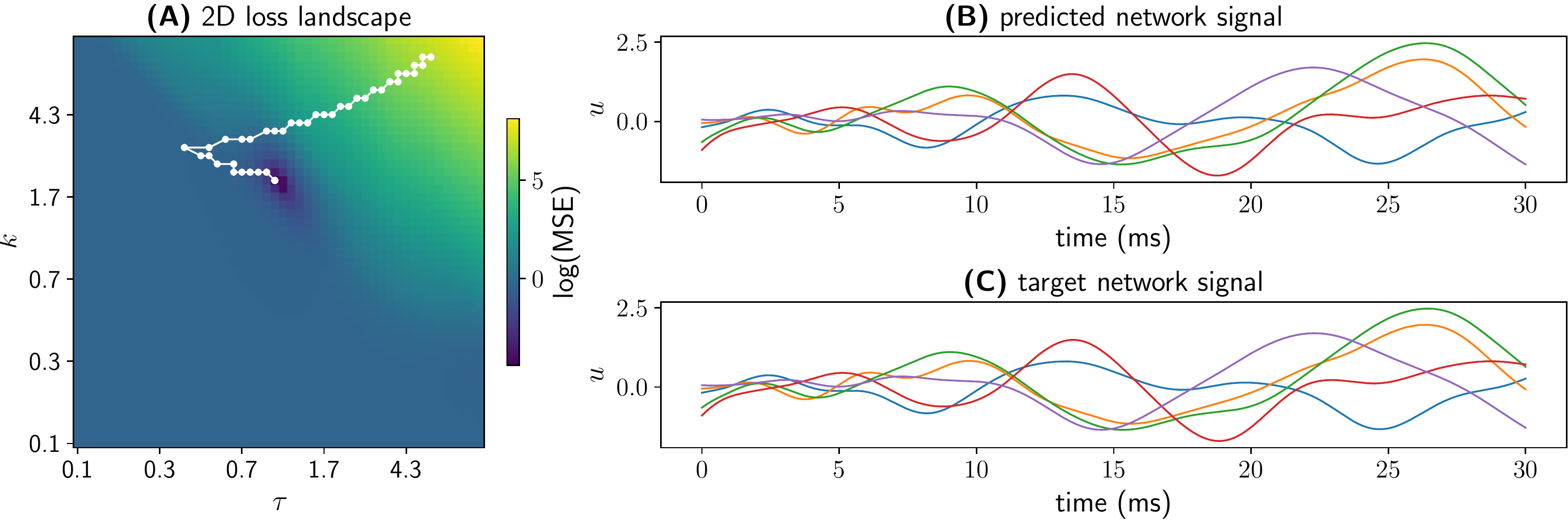}
    \caption{Comparison of the dynamics of the target leaky integrator model and the fitted leaky integrator model. \textbf{(A)} Logarithm of the mean-squared error (color-coded), depicted over the search range of the two parameters that were optimized: $\kappa$ and $\tau$. The white trace shows the steps taken by the optimizer from its initialization point to the global minimum. \textbf{(B and C)} Rate signals of all $N$ LI units over time of the fitted network and the target network, respectively.}
    \label{fig:li_opt}
\end{figure}

The results of the parameter optimization are depicted in Fig.\ref{fig:li_opt}. 
As can be seen, the optimization algorithm succeeded in finding values of the parameters $\tau$ and $k$ for which the network reproduces the target dynamics of the $5$ leaky integrators.

In conclusion, we successfully used the \textit{PyTorch} equations generated by \textit{PyRates} to run parameter optimizations via \textit{RectiPy}.

\section{Discussion}

In this work, we provided an overview of features and use cases of the dynamical systems modeling software \textit{PyRates}.
We introduced the structure of the software and described how \textit{PyRates} (i) supports the implementation of dynamical system models based on ordinary or delayed differential equations, (ii) provides access to various simulation backends such as \textit{NumPy} or \textit{Fortran}, and (iii) allows for the generation of backend-specific equation files.
The latter can be used to interface various dynamical system analysis tools, such as the ones listed in Tab.~\ref{tab:ds_tools}.
This way, model implementations in \textit{PyRates} serve as a starting point for flexible dynamical system analysis workflows that allow users to choose between a large variety of analysis tools and methods at each step of the workflow.
Our examples demonstrated this by using \textit{PyRates} to (i) perform numerical integration of an ODE system via \lstinline{scipy.integrate.solve_ivp} (see section \ref{sec:vdp}), (ii) generate the Fortran files to run bifurcation analysis in \textit{Auto-07p} (see section \ref{sec:qif}), and (iii) generate the equations for a \textit{PyTorch} compute graph to perform parameter optimization via \textit{RectiPy} (see section \ref{sec:li}).
\textit{PyRates} thus contributes to the minimization of dynamical system model implementation errors and to the setup of efficient, flexible, and reproducible dynamical system analysis workflows \citep{leveque_reproducible_2012,topalidou_long_2015,piccolo_tools_2016}.


\subsection{Limitations}

The main limitation of \textit{PyRates} is the family of dynamical system models it supports.
Currently, \textit{PyRates} provides support for ODE and DDE systems, and state variables of these systems can be real or complex-valued.
Examples of each of these differential equation types can be found in the use example section at \url{https://pyrates.readthedocs.io/en/latest/}.
\textit{PyRates} does not currently support partial differential equations (PDEs), which involve derivatives in multiple variables and can be used to model dynamical systems in continuous time and space, and stochastic differential equations (SDEs), which are typically used to model inherent stochastic fluctuations of dynamical processes.
Both types of differential equation systems have been widely used in dynamical system modeling \citep{hutt_synergetics_2020}.
In neuroscience, for instance, PDEs have been applied in the context of neural field models \citep{deco_dynamic_2008,coombes_large-scale_2010}.
A number of dynamical system analysis libraries currently supported by \textit{PyRates} such as \textit{SciPy} \citep{virtanen_scipy_2020} or \textit{DifferentialEquations.jl} \citep{rackauckas_differentialequationsjl_2017} provide algorithms for the numerical integration of PDEs and SDEs. 
Thus, adding support for these types of differential equations would be a useful extension to the currently supported list of dynamical system models.

Another limitation of \textit{PyRates} is its inability to define specific events that may occur during the numerical integration of a differential equation system.
An example of such an event is the membrane potential of a neuron crossing a certain threshold and eliciting a spike, which can be modeled as a singular event in time \citep{dayan_theoretical_2001}.
Events like this introduce discontinuities to the differential equation system, which are not currently supported by \textit{PyRates}.
However, although \textit{PyRates} does not support event definition in general, the \textit{PyRates} extension \textit{RectiPy} introduced in Sec\ref{sec:li} allows for the definition of spike conditions for the specific case of spiking neural networks.
\textit{RectiPy} provides support for numerical simulations and parameter optimization, and supports the use of both rate neurons and spiking neurons.


\subsection{Outlook}

It is important to note that the limitations of \textit{PyRates} are not inherent limitations that cannot be overcome by the software; rather, they are areas where the software has not yet been extended.
Due to the highly modular structure and open-source nature of \textit{PyRates}, such extensions can readily be implemented.
For example, adding another backend to \textit{PyRates} can be done without any changes to the frontend, whereas added support for SDEs would mostly involve changes to the frontend.
Additionally, some of the limitations outlined above can be addressed by using additional software packages that extend \textit{PyRates} with specific functionalities.
We have shown here that packages that extend \textit{PyRates} can simply be built by employing \textit{PyRates} as a model definition interface and instructing it to generate the output files required for a specific extension. 
We have demonstrated that by generating \textit{Fortran} files for \textit{PyCoBi} and \textit{PyTorch} files for \textit{RectiPy}, which are software packages for bifurcation analysis and artificial neural network training, respectively.

In summary, \textit{PyRates} already supports a large family of dynamical system systems and backends, and is designed be easily extendable in the future.
This makes it a versatile dynamical system model definition language and code-generation tool that provides access to a wide variety of dynamical system analysis methods and allows for sharing models without being tied to specific programming languages or analysis tools.

\section{Acknowledgments}

We would like to thank the Michael J. Fox Foundation for their support of R.G. via the Aligning Science Across Parkinson’s grant (ASAP-020551) awarded to A.K.

\bibliography{references}










\end{document}